\documentclass[a4paper,11pt]{article}
\usepackage{cite}
\usepackage{layaureo}
\usepackage[]{graphicx}
\usepackage[]{amsmath}
\usepackage{amssymb,stmaryrd}
\usepackage{listings}
\usepackage{multirow,array}
\usepackage{url}
\usepackage{syntax}
\makeatletter
\renewcommand{\ulitleft}{\normalfont\bf\ttfamily\syn@ttspace\frenchspacing}
\renewcommand{\ulitright}{}
\renewcommand{\litleft}{`\bgroup\ulitleft}
\renewcommand{\litright}{\ulitright\egroup'}
\makeatother
\usepackage{mathtools}
\usepackage{fixltx2e}
\usepackage{tikz}
\usetikzlibrary{automata}
\usetikzlibrary{shapes.geometric}
\usetikzlibrary{arrows}
\usepackage{tikz-qtree}
\usepackage{paralist}
\usepackage{xspace,url}
\usepackage[caption=false,font=footnotesize]{subfig}

\lstdefinelanguage{simplec}[]{C}{morekeywords={true,false,then,endif,endwhile,begin,end},
}
\newcommand{\ft}[1]{{\footnotesize #1}}
\newcommand{\upd}{\ensuremath{\mathit{upd}}\xspace} 

\newtheorem{mydef}{Definition}
\newtheorem{myprop}{Proposition}
\newcommand{\sidecar}{SiDECAR\xspace}
\newcommand{\stt}[1]{\begin{footnotesize}\texttt{#1}\end{footnotesize}}
\usepackage[charter]{mathdesign}
\usepackage[scaled]{beramono}

\begin{document}
\title{A Syntactic-Semantic Approach to\\ Incremental Verification}

\author{Domenico Bianculli\\
SnT Centre, University of Luxembourg, Luxembourg, Luxembourg\\
\texttt{domenico.bianculli@uni.lu}
\and
{Antonio Filieri}\\
{Institute of Software Technology}, University of Stuttgart, Stuttgart, Germany\\
\texttt{antonio.filieri@informatik.uni-stuttgart.de}
\and
{Carlo Ghezzi}\\ {DEEPSE group - DEI}, Politecnico di Milano, Milano, Italy\\
\texttt{ghezzi@elet.polimi.it}
\and
{Dino Mandrioli}\\{DEEPSE group - DEI}, Politecnico di Milano, Milano, Italy\\
\texttt{mandrioli@elet.polimi.it}
}

\maketitle
\begin{abstract}
Software verification of evolving systems is challenging mainstream
methodologies and tools.  Formal verification techniques often
conflict with the time constraints imposed by change management
practices for evolving systems. Since changes in these systems are
often local to restricted parts, an incremental verification approach
could be beneficial.

This paper introduces SiDECAR, a general framework for the definition
of verification procedures, which are made incremental by the
framework itself.  Verification procedures are driven by the syntactic
structure (defined by a grammar) of the system and encoded as semantic
attributes associated with the grammar.  Incrementality is achieved by
coupling the evaluation of semantic attributes with an incremental
parsing technique.

We show the application of SiDECAR to the
definition of two verification procedures: probabilistic verification
of reliability requirements and verification of safety properties.

\end{abstract}

\maketitle
\newpage
\section{Introduction}
\label{secIntroduction}

Software evolution is a well-known phenomenon in
software engineering. Software may evolve
because of a change in the requirements or in the domain assumptions,
leading to the development and deployment of many new versions of
the software. This phenomenon is taken to extremes by new kinds of
software, called \emph{open-world software}~\cite{Baresi2006},
built by composing heterogeneous, third-party components, whose
behavior and interactions cannot be fully controlled or
predicted. This software is required to react to changes in its
environment, 
by bringing verification to run time~\cite{calinescu12:self} and (self-) adapting its behavior while it is executing.

Incremental verification has been suggested as a possible approach to dealing
with evolving software~\cite{Sistla:1996:HIM:242224.242384}.
An incremental verification approach tries to reuse as
much as possible the results of a previous verification step, and
accommodates within the verification procedure---possibly in a
``smart'' way---the changes occurring in the new version. By
avoiding re-executing the verification process from scratch,
incremental verification may considerably reduce the verification
time. This may be appealing for adoption within agile development
processes. Moreover, incremental verification 
may speed up change management, which may be subject to 
severe time constraints, especially if it needs to
be performed at run time, to support dynamic self-adaptation.
	
This paper proposes SiDECAR (Syntax-DrivEn inCrementAl veRification),
a general framework to define verification procedures, which are
automatically enhanced with incrementality by the framework itself.
The framework follows a syntactic-semantic approach, since it assumes
that the software artifact to be verified has a syntactic structure
described by a formal grammar, and that the verification procedure is
encoded as synthesis of semantic attributes~\cite{knuth1968}, associated with the
grammar and evaluated by traversing the syntax tree of
the artifact. 
 We based the framework on operator precedence
grammars~\cite{Floyd1963},  which allow for re-parsing, and hence semantic
re-analysis, to be confined within 
 an inner
portion of the input that encloses the changed part.
This property is the key for an efficient incremental 
verification procedure:
since the
verification procedure is encoded within
attributes, their evaluation proceeds incrementally, hand-in-hand with
parsing.

The main contributions of the
paper are:
\begin{inparaenum}[i)]
  \item the definition of a methodological approach for incremental
    syntactic-semantic verification procedures (\sidecar);
  \item the application of \sidecar to 
the definition of two verification procedures:
 probabilistic verification
of reliability requirements and verification of safety properties.
\end{inparaenum}
Indeed, the goal of the paper is to present the general framework, which can
be used to define incremental verification procedures. The two examples are
provided to show the generality and versatility of the approach. 

The rest of the paper is structured as follows.
Section~\ref{sec:background} introduces some background concepts on
operator precedence grammars and attribute grammars. Section~\ref{sec:sidecar} shows
how \sidecar exploits operator precedence grammars to support
syntactic-semantic incremental verification. In
section~\ref{sec:usingsidecar} we show \sidecar at work, by 
presenting the two examples. In section~\ref{sec:discussion} we
discuss the application of the methodology supported by \sidecar.
Section~\ref{secRelatedWork}
presents related work. Section~\ref{secConclusions} provides some
concluding remarks.

\section{Background}
\label{sec:background}
Hereafter we briefly recall the definitions of operator precedence grammars and attribute grammars.
For more information on formal languages and grammars, we refer the reader to~\cite{grune08:parsin-techn} and~\cite{crespi-reghizzi2010:operator-preced}.

\subsection{Operator precedence Grammars}

We start by recalling the definition of a \emph{context-free (CF)} grammar $G$ as a tuple $G=\langle V_N,V_T,P,S \rangle$, where $V_N$ is a finite set of non-terminal symbols; $V_T$ is a finite set of terminal symbols, disjoint from $V_N$; $P \subseteq V_N \times (V_N \cup V_T)^*$ is a relation whose elements represent the rules of the grammar; $S \in V_N$ is the axiom or start symbol. 
We use the following naming convention, unless otherwise specified:
non-terminal symbols are enclosed within chevrons, such as \synt{A}; terminal ones are enclosed within single quotes, such as \lit{+} or are denoted by lowercase letters at the beginning of the alphabet $(a,b,c, \ldots)$; lowercase letters at the end of the alphabet $(u, v, x, \ldots )$ denote terminal strings; $\varepsilon$ denotes the empty string.
For the notions of \emph{immediate derivation} ($\Rightarrow$), \emph{derivation} $(\stackrel{*}{\Rightarrow})$, and the language $L(G)$ generated by a grammar $G$ please refer to the standard literature, e.g.,~\cite{grune08:parsin-techn}.

A rule is in \emph{operator form} if its right hand side (rhs) has no adjacent non-terminals; an \emph{operator grammar (OG)} contains only rules in operator form.

\begin{figure}[tb]
\begin{small}
  \centering
  \subfloat[]{    \begin{tabular}[b]{p{.5em}>{$}p{.5em}<{$}p{38mm}}
      \synt{S} & \Coloneqq & \synt{A} $\mid$ \synt{B}\\
      \synt{A} & \Coloneqq & \synt{A} \lit{+} \synt{B} $\mid$ \synt{B} \lit{+} \synt{B}\\
      \synt{B} & \Coloneqq & \synt{B} \lit{*} \lit{n} $\mid$ \lit{n} \\
    \end{tabular}
  \label{fig:cfg}
  }  \quad
  \subfloat[]{
\begin{tabular}[b]{p{0.5em}@{\hspace{10pt}}|>{$}p{0.5em}<{$}>{$}p{0.5em}<{$}>{$}p{0.5em}<{$}}
     & \lit{n}        & \lit{*}        & \lit{+} \\ \hline
   \lit{n} &          & \gtrdot  & \gtrdot \\
   \lit{*} & \doteq   &          & \\
   \lit{+} & \lessdot & \lessdot & \gtrdot \\
    \end{tabular}
  \label{fig:opm}
  }                      \end{small}
\caption{Example of an operator grammar (`\texttt{\textbf{n}}' stands for any natural number) and its operator precedence  matrix}
\label{fig:grammars}

\end{figure}

\emph{Operator precedence grammars (OPGs)}~\cite{Floyd1963} are defined starting from operator grammars by means of binary relations on $V_T$ named \emph{precedence}.  Given two terminals, the precedence relations between them can be of three types: \emph{equal-precedence} ($\doteq$), \emph{takes-precedence} ($\gtrdot$), and \emph{yields-precedence} ($\lessdot$). The meaning of precedence relations is analogous to the one between arithmetic operators and is the basic driver of deterministic parsing for these grammars. Precedence relations can be computed in an  automatic way for any operator grammar.
We represent the precedence relations in 
a $V_T \times V_T$ matrix, named \emph{operator precedence matrix (OPM)}. An entry $m_{a,b}$ of an OPM represents the set of operator precedence relations holding between terminals $a$ and $b$. For example, Fig.~\ref{fig:opm} shows the OPM for the grammar of arithmetic expressions in Fig.~\ref{fig:cfg}.
Precedence relations 
have to be neither reflexive, nor symmetric, nor transitive, nor total.
If an entry $m_{a,b}$ of an OPM $M$ is empty, the occurrence of the terminal $a$ followed by the terminal $b$ represents a malformed input, which cannot be generated by the grammar.
\begin{mydef}[Operator Precedence Grammars]
  An OG $G$ is an OPG grammar if and only if its OPM is a conflict-free matrix, i.e., for each $a,b \in V_T, |m_{a,b}| \leq 1$.
\end{mydef}
\begin{mydef}[Fischer Normal Form, from~\cite{crespi-reghizzi2010:operator-preced}]
An OPG is in Fischer Normal Form (FNF) if it is invertible, the axiom \synt{S} does not occur in the right-hand side (rhs) of any rule, no empty rule exists except possibly $\synt{S} \Rightarrow \varepsilon$, the other rules having \synt{S} as left-hand side (lhs) are renaming, and no other renaming rules exist.
\end{mydef}

The grammar of Fig.~\ref{fig:cfg} is in FNF. In the sequel, we assume, without loss of generality, that OPGs are in FNF. Also, as is customary in the parsing of OPGs, the input strings are implicitly enclosed between
two \lit{\#} special characters, such that \lit{\#} yields precedence to any other character
and any character takes precedence over \lit{\#}. The key feature of OPG parsing is that a sequence of terminal characters enclosed within a pair $\lessdot$ $\gtrdot$ and separated by $\doteq$ uniquely determines a rhs to be replaced, with a shift-reduce algorithm, by the corresponding lhs. Notice that in the parsing of these grammars non-terminals are ``transparent'', i.e., they are not considered for the computation of the precedence relations.
For instance, consider the syntax tree of Fig.~\ref{fig-expressionSyntaxTree01}
generated by the grammar of Fig.~\ref{fig:cfg}: the leaf \lit{6} is preceded by \lit{+}
and followed by \lit{*}. Because \lit{+} $\lessdot$ \lit{6}  $\gtrdot$ \lit{*}, \lit{6} is reduced to \synt{B}. Similarly, in a further step we have \lit{+} $\lessdot$ \synt{B} \lit{*} $\doteq$ \lit{7} $\gtrdot$ \lit{*} and we apply the reduction \synt{B} $\Rightarrow$ \synt{B} \lit{*} \lit{7} (notice that non-terminal \synt{B} is ``transparent'') and so on.

\subsection{Attribute Grammars}

\emph{Attribute Grammars (AGs)} have been proposed by Knuth as a way to express the semantics of programming languages \cite{knuth1968}. AGs extend CF grammars by associating \emph{attributes} and semantic functions to the rules of a CF grammar; attributes define the ``meaning'' of the corresponding nodes in the syntax tree.
In this paper we consider only
\emph{synthesized} attributes, which characterize an information flow
from the children nodes (of a syntax tree) to their parents; more general attribute schemas do not add semantic power~\cite{knuth1968}.

An AG is 
obtained from a CF grammar $G$ by adding a finite set of attributes $\mathit{SYN}$ and a set $\mathit{SF}$ of semantic functions. Each symbol $X \in V_N$ has a  set of (synthesized) attributes $\mathit{SYN}(X)$; $\mathit{SYN} = \bigcup_{X \in V_N}  \mathit{SYN}(X)$. 
We use the symbol $\alpha$ to denote a generic element of $\mathit{SYN}$; we assume that each $\alpha$ takes values in a corresponding domain $T_\alpha$.
The set $\mathit{SF}$ consists of functions, each of them associated with a rule $p$ in $P$. 
For each attribute $\alpha$ of the lhs of $p$, a function $f_{p\alpha} \in \mathit{SF}$
synthesizes the value of $\alpha$ based on the attributes of the non-terminals in the rhs of $p$.
For example, the grammar in Fig.~\ref{fig:cfg} can be extended to an attribute grammar that computes the value of an expression. 
All nodes have only one attribute called \textit{value}, with $T_{\mathit{value}}=\mathbb{N}$.
The set of semantic functions $\mathit{SF}$ is defined as in Fig.~\ref{fig:ag}, where semantic functions are enclosed in braces next to each rule:
\begin{figure}[tb]
\begin{small}
\begin{tabular}[]{p{.5em}>{$\;}p{.5em}<{$}l@{\hspace{5pt}}>{$\{}l<{$}@{\hspace{2pt}}>{$}c<{$}@{\hspace{2pt}}>{$}l<{$}}
  \synt{S} & \Coloneqq & \synt{A} &  \mathit{value}(\synt{S}) & = &  \mathit{value}(\synt{A})\}\\
  \synt{S} & \Coloneqq & \synt{B} &  \mathit{value}(\synt{S}) & = &  \mathit{value}(\synt{B})\}\\
      \synt{A\textsubscript{0}} & \Coloneqq & \synt{A\textsubscript{1}} \lit{+} \synt{B} & \mathit{value}(\synt{A\textsubscript{0}})& =  &  \mathit{value}(\synt{A\textsubscript{1}}) + \mathit{value}(\synt{B})\}\\
      \synt{A} & \Coloneqq & \synt{B\textsubscript{1}} \lit{+} \synt{B\textsubscript{2}} & \mathit{value}(\synt{A})& =  &  \mathit{value}(\synt{B\textsubscript{1}}) + \mathit{value}(\synt{B\textsubscript{2}})\}\\
      \synt{B\textsubscript{0}} & \Coloneqq & \synt{B\textsubscript{1}} \lit{*} \lit{n} & \mathit{value}(\synt{B\textsubscript{0}})& =  &  \mathit{value}(\synt{B\textsubscript{1}}) * \mathit{eval}(\lit{n})\}\\
      \synt{B}   & \Coloneqq & \lit{n}     & \mathit{value}(\synt{B})& =  &  \mathit{eval}(\lit{n})\} \\
\end{tabular}
\end{small}
\caption{Example of attribute grammar}
\label{fig:ag}
\end{figure}
The $+$ and $*$ operators appearing within braces correspond, respectively, to the standard operations of arithmetic addition and multiplication, and $eval(\cdot)$ evaluates its input as a number. Notice also that, within a rule, different occurrences of the same grammar symbol are denoted by distinct subscripts.

\section{\sidecar and Syntactic-semantic Incrementality}
\label{sec:sidecar}

\sidecar exploits a syntactic-semantic approach to define verification
procedures that are encoded as semantic functions associated with an
attribute grammar. In this section we show how OPGs,
equipped with a suitable attribute schema, can support incrementality
in such verification procedures in a natural and efficient way.

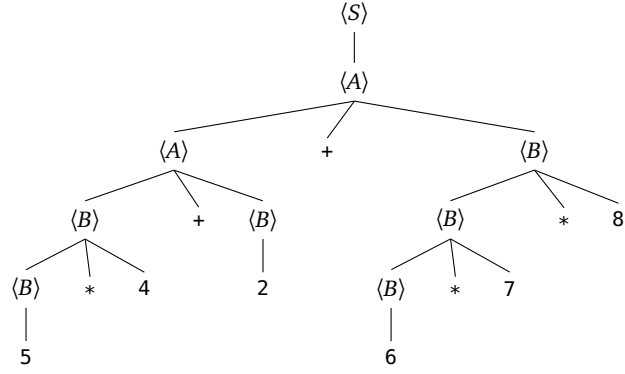
\begin{figure}[tb]
\begin{footnotesize}
\centering
\begin{tikzpicture}[level distance=26pt,sibling distance=10pt]
\Tree [
 .\synt{S} [
 .\synt{A}  [ .\synt{A} [ .\synt{B}  [.\synt{B} \texttt{5} ] \texttt{*} \texttt{4} ]  \texttt{+} [.\synt{B} \texttt{2} ]  ]
 \texttt{+} 
 [ .\synt{B} [ .\synt{B} [.\synt{B} \texttt{6} ] \texttt{*} \texttt{7} ] \texttt{*} \texttt{8}  ] 
]
]
\end{tikzpicture}

\caption{Abstract syntax tree of the expression `\texttt{\textbf{{5*4+2+6*7*8}}}'}
\label{fig-expressionSyntaxTree01}
\end{footnotesize}
\end{figure}

\subsection{The Locality Property and Syntactic Incrementality}
\label{secSyntacticIncrementality}

The main reason for the choice of OPGs is that, unlike 
more commonly 
used grammars that support deterministic parsing, they
enjoy the \textit{locality property}, i.e., the possibility of
starting the parsing from any arbitrary point of the sentence to be
analyzed, independent of the context within which the sentence is
located. In fact for OPGs the following proposition holds.
\begin{myprop}
  If $a\synt{A}b \stackrel{*}{\Rightarrow} asb$, then, for every $t,
  u$, $\synt{S} \stackrel{*}{\Rightarrow} tasbu $ if{f} $\synt{S}
  \stackrel{*}{\Rightarrow} ta\synt{A}bu \stackrel{*}{\Rightarrow}
  tasbu$. As a consequence, if $s$ is replaced by $v$ in the context
  $\llbracket ta, bu \rrbracket$, and $a\synt{A}b \stackrel{*}{\Rightarrow}
  avb$, then $\synt{S} \stackrel{*}{\Rightarrow} ta\synt{A}bu
  \stackrel{*}{\Rightarrow} tavbu$, and (re)parsing of $tavbu$ can be
  stopped at $a\synt{A}b \stackrel{*}{\Rightarrow} avb$.
\end{myprop}

Hence, if we build---by means of a bottom-up parser---the derivation
$a\synt{A}b \stackrel{*}{\Rightarrow} avb$, we say that a
\emph{matching condition} with the previous derivation $a\synt{A}b
\stackrel{*}{\Rightarrow} asb$ is satisfied and we can replace the old
subtree rooted in \synt{A} with the new one, independently of the global
context $\llbracket ta, bu\rrbracket$ (only the local context $\llbracket a, b
\rrbracket$ matters for the incremental parsing).

For instance, consider the string and syntax tree of
Fig.~\ref{fig-expressionSyntaxTree01}. Assume that the expression is
modified by replacing the term \lit{6*7*8} with \lit{7*8}.  The corresponding
new subtree can clearly be built independently within the context
$\llbracket \text{\lit{+}}, \text{\lit{\#}} \rrbracket$.  The matching condition is
satisfied by $\lit{+} \synt{B} \lit{\#} \stackrel{*}{\Rightarrow}
\lit{+} \lit{6} \lit{*} \lit{7} \lit{*} \lit{8} \lit{\#}$ and $\lit{+}
\synt{B} \lit{\#} \stackrel{*}{\Rightarrow} \lit{+} \lit{7} \lit{*}
\lit{8} \lit{\#}$; thus the new subtree can replace the original one
without affecting the remaining part of the global tree.  If, instead,
we replace the second \lit{+} by a \lit{*}, the affected portion of
syntax tree would be larger and more re-parsing would be
necessary\footnote{  Some further optimization could be applied by integrating
  the matching condition with techniques adopted
  in~\cite{Ghezzi1979} (not reported here for brevity).}.

In general, the incremental parsing algorithm, for any replacement of
a string $w$ by a string $w^\prime$ in the context $\llbracket t, u
\rrbracket$, automatically builds the minimal ``sub-context'' $\llbracket
t_1, u_1 \rrbracket$ such that for some \synt{A}, $a\synt{A}b
\stackrel{*}{\Rightarrow} at_1wu_1b$ and $a\synt{A}b
\stackrel{*}{\Rightarrow} at_1w^\prime u_1b$.

The locality property\footnote{The locality property has also been
 shown to support an efficient parallel parsing
  technique~\cite{barenghi12}, which is not further exploited here.} has a price in terms of generative power.
For example, the LR grammars traditionally used to describe and parse
programming languages do not enjoy it. However they can generate all
the deterministic languages. OPGs cannot; this
limitation, however, is more of theoretical interest than of real practical
impact. Large parts of the grammars of many computer languages are
operator precedence~\cite[p. 271]{grune08:parsin-techn}; 
a complete OPG is available for
Prolog~\cite{bosschere1996:an-operator-pre}. 
Moreover, in many practical cases one can obtain an OPG by minor
adjustments to a non operator-precedence grammar~\cite{Floyd1963}.

In the current \sidecar prototype, we developed an incremental parser
for OPGs that exhibits the following features: linear
complexity in the length of the string, in case of parsing from
scratch; linear complexity in the size of the \emph{modified
  subtree(s)}, in case of incremental parsing; $O(1)$ complexity of
the matching condition test.

\begin{figure}[tb]
  \centering
\begin{tikzpicture}
\begin{scope}
 \tikzset{every node/.style={isosceles triangle, draw,anchor=apex, shape border rotate=90, isosceles triangle stretches}}
 \node[minimum height=35mm,minimum width=8cm] (out) at (4,4) {};
 \node[minimum height=15mm,minimum width=2.5cm] (in) at (4.2,2) {};
 \node[minimum height=7.5mm,minimum width=10mm] (in2) at (4.5,1.25) {};
\end{scope}
 \node (S) at (4,4.3) {$\alpha_S$};
 \node (M) at (3.75,2.2) {$\alpha_M$};
 \node (N) at (4.2,1.3) {$\alpha_N$};
 \node (K) at (4.5,2.75) {$\alpha_K$};
 \node (P) at (2.5,1.5) {$\alpha_P$};
 \node (Q) at (5.5,1.5) {$\alpha_Q$};
 \node (t) at (4,0) {$xw^\prime z$};
 \draw[-latex] (t.east) -- (in.right corner);
 \draw[-latex] (t.west) -- (in.left corner);
 \draw[densely dashed] (in2.apex) -- (in.apex);
 \draw[densely dashed] (in.apex) -- (K);
 \draw[densely dashed] (K) -- (out.apex);
 \draw[o-] (P) -- (in.apex);
 \draw[o-] (Q) -- (in.apex);
\end{tikzpicture}

\caption{Incremental evaluation of semantic attributes}
\label{fig-attributes01}
\end{figure}
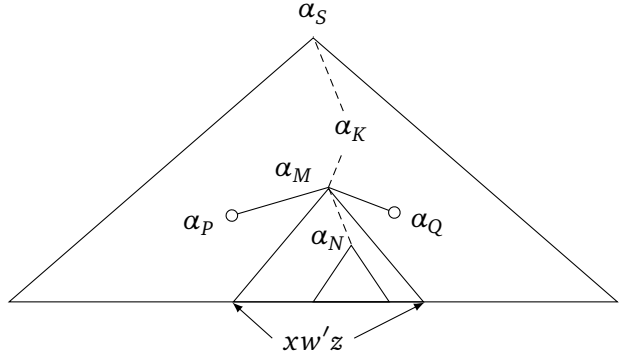

\subsection{Semantic Incrementality}
\label{secSemanticIncrementality}

In a bottom-up parser, semantic actions are performed during a
reduction. This allows the re-computation of semantic attributes after
a change
to proceed hand-in-hand with the re-parsing of the modified substring.
Suppose that, after replacing substring $w$ with $w^\prime$,
incremental re-parsing builds a derivation $\synt{N}
\stackrel{*}{\Rightarrow} x w^\prime z$, with the same non-terminal
\synt{N} as in $\synt{N} \stackrel{*}{\Rightarrow} xwz$, so that the
matching condition is verified.
Assume also that \synt{N} has an attribute $\alpha_N$. 
Two situations may occur related to the computation of $\alpha_N$:
\begin{asparaenum}
\item The $\alpha_N$ attribute associated with the new subtree rooted
  in \synt{N} has the same value as before the change. In this case,
  all the remaining attributes in the rest of the tree will not be
  affected, and no further analysis is needed.
\item The new value of $\alpha_N$ is different from the one it had
  before the change.  In this case (see Fig.~\ref{fig-attributes01})
  only the attributes on the path from \synt{N} to the root \synt{S}
  (e.g., $\alpha_M, \alpha_K, \alpha_S$) may change and  in such case
  they need to
  be recomputed. The values of the other attributes not on the path
  from \synt{N} to the root (e.g., $\alpha_P$ and $\alpha_Q$) do not
  change: there is no need to recompute them.
\end{asparaenum}

\section{\sidecar at work}
\label{sec:usingsidecar}
Using \sidecar requires to define 1) an OPG for the programming language
one wants to support and 2) the associated attribute grammar schema
corresponding to the verification procedures that one wants to
implement. In this section we use programs written in the \emph{Mini} language,
whose OPG is shown in Fig.~\ref{fig:simplecgrammar}.
It is a minimalistic language that 
includes the major constructs of structured
programming.
  For the sake
of readability and to reduce the complexity of attribute schemas,
\emph{Mini} programs support only (global) boolean variables and
boolean functions (with no input parameters). These
assumptions can be relaxed, with no impact on the applicability of
our approach.

In the rest of this section we demonstrate  the generality of the \sidecar
framework by means of two examples of incremental verification. The
former one
(Section~\ref{secReliabilityAnalysis}) reports on probabilistic
verification of reliability properties of programs that compose
possibly faulty functions. The latter
(Section~\ref{sec:reachability-analysis}) reports on verification of
safety properties of programs. We chose two simple, but rather
diverse examples to demonstrate \sidecar 's versatility as a
general framework.
For
space reasons and for the sake of readability, we adopt a
straightforward encoding of these verification procedures and make several
simplifying assumptions. We deliberately omit all optimizations and
heuristics that would improve the verification, which are adopted by
state-of-the-art tools. Nevertheless these could be accommodated in
\sidecar through richer (and more complex) attributes.

\begin{figure}[tb]
\begin{small}
  \centering
  \begin{grammar}
 <S> ::= `begin' <stmtlist> `end'

 <stmtlist> ::= <stmt> `;' <stmtlist> 
  \alt  <stmt> `;'

  <stmt> ::= <function-id> `(' `)'
  \alt <var-id> `:=' `true'
  \alt <var-id> `:=' `false'
  \alt <var-id> `:=' <function-id> `(' `)' 
  \alt `if'  <cond> `then' <stmtlist> `else' <stmtlist> `endif'
  \alt `while' <cond> `do' <stmtlist> `endwhile'

 <var-id> ::= \ldots

  <function-id> ::= \ldots

 <cond> ::= \ldots
\end{grammar}
\end{small}
\caption{The grammar of the \textit{Mini} language}
  \label{fig:simplecgrammar}
\end{figure}

\begin{figure}[tb]
  \centering
\begin{tikzpicture}[scale=0.7]
\Tree [ 
 .{\synt{S} \ft{0}} 
 [.{\synt{stmlist} \ft{1}}
 [.{\synt{stmt} \ft{2}}
                       [.{\synt{function-id} \ft{3}}
                       {\texttt{opA()} \ft{4}} ] 
 ]
 [.{\synt{stmlist} \ft{5}} 
  [.{\synt{stmt} \ft{6}}
    [
     .{\synt{var-id} \ft{7}} {\texttt{x} \ft{8}}
    ]
     {\texttt{true} \ft{9}}
  ]
  [.{\synt{stmlist} \ft{10}}
   [
     .{\synt{stmt} \ft{11}}
     [.{\synt{cond} \ft{12}} {\texttt{x==true} \ft{13}}
     ]
     [
      .{\synt{stmlist} \ft{14}} [.{\synt{stmt} \ft{15}}
                       [.{\synt{function-id} \ft{16}}
                       {\texttt{opB()} \ft{17}} ] 
                      ]
     ]
     [
      .{\synt{stmlist} \ft{18}} [.{\synt{stmt} \ft{19}}
                       [.{\synt{function-id} \ft{20}}
                       {\texttt{opA()} \ft{21}} ] 
                      ]
    ]
  ]
 ]
]
]
]
\begin{scope}[xshift=-1cm,yshift=-6.2cm]
\draw[densely dashed] (-2,.5) rectangle (2,-2.25);
\Tree [.{\synt{stmt} \ft{6}}
    [
     .{\synt{var-id} \ft{7}} {\texttt{x} \ft{8}}
    ]
     {\texttt{false} \ft{9}}
  ]
\end{scope}

\end{tikzpicture}
  \caption{The syntax tree of version 1 of the example program;
the subtree in the box shows the difference (node 9) in the syntax tree of version 2}
  \label{fig:trees}
\end{figure}
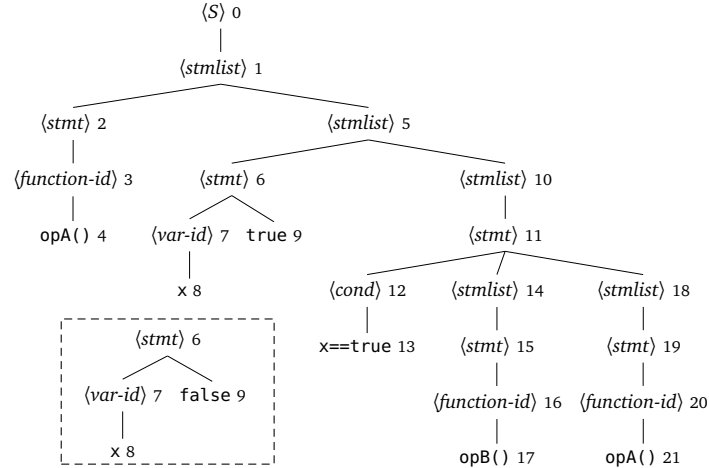

To show the benefits of incrementality, 
for each of the verification procedures defined in the next
subsections, we  analyze two versions of the same example
program (shown in
Fig.~\ref{fig:program-versions}),  which differ in the assignment at
line~\ref{change}, which determines the execution of the subsequent
\emph{if} statement, with implications on the results of the two
analyses.  Figure~\ref{fig:trees} depicts the syntax tree of version~1
of the program, as well as the subtree that is different in version~2;
nodes of the tree have been numbered for quick reference.

\begin{small}
\newsavebox{\versiona}
\begin{lrbox}{\versiona}
\begin{minipage}[T]{0.45\linewidth}
\begin{lstlisting}
begin
 opA();
 x := true; (*@\label{change}@*) 
 if (x==true) 
  then opB();
  else opA();
 endif;
end
\end{lstlisting}
\end{minipage}
\end{lrbox}

\newsavebox{\versionb}
\begin{lrbox}{\versionb}
\begin{minipage}[T]{0.45\linewidth}
\begin{lstlisting}
begin
 opA();
 x := false;
 if (x==true) 
  then opB();
  else opA();
 endif;
end
\end{lstlisting}
\end{minipage}
\end{lrbox}

\begin{figure}[b]
\centering
\subfloat[Version 1]{\usebox{\versiona}} \hfill\subfloat[Version 2]{\usebox{\versionb}}
  \caption{The two versions of the example program}
  \label{fig:program-versions}
\end{figure}

\end{small}

The next two subsections describe in detail the two analyses and
their corresponding attribute schemas. Before presenting them, here
we introduce some useful notations. Given a \emph{Mini} program $P$,
$F_P$ is the set of functions and $V_P$ the
set of variables defined within $P$; $E_P$ is the set of
boolean expressions that can appear as the condition of an \emph{if} or a
\emph{while} statement in $P$.  An expression $e \in E_P$ is either a
combination of boolean predicates on program variables or a
placeholder predicate labeled $\ast$. 
Hereafter, we drop the subscript $P$ in $F_P$, $V_P$, and $E_P$
whenever the program is clear from the context.

\subsection{Probabilistic Verification of Reliability Requirements}
\label{secReliabilityAnalysis}
In this section we show how to apply \sidecar to perform probabilistic
verification of reliability
requirements of \emph{Mini} programs.  Reliability is a ``user-oriented''
property~\cite{Cheung1980}; in other words, a software may be more or
less reliable depending on its use. If user inputs do not activate a
fault, a failure may never occur even in a software containing
defects~\cite{Avizienis2004ix}; on the other hand, users may stress a
faulty component, leading to a high frequency of failure events.  Here
we consider reliability as the probability of
successfully accomplishing an assigned task, when requested. 

We observe that the verification problem presented here for
\emph{Mini} can be viewed as a high-level abstraction of a similar
verification problem for service compositions in the context of
service-oriented architectures, since the call to possibly faulty
functions mimics the call to third-party services.

Most of the current approaches for verification of reliability requirements use
probabilistic model
checking~\cite{pham2006system,immonen2008survey}. Software systems are
translated into stochastic models, such as Discrete Time Markov Chains
(DTMCs), which are suitable to represent usage profiles and failure probabilities.
A DTMC is essentially a finite state automaton where states abstract
the program execution state, such as the execution of a task or the
occurrence of a failure, and the transitions among states are defined
through a probabilistic distribution.  DTMCs can be analyzed with
probabilistic model checkers such as PRISM~\cite{prismsymbolic} and
MRMC~\cite{mrmc}.

To model the probabilistic verification problem in \sidecar, first we
assume that each function $f \in F$ has a probability
$\mathit{Pr}_S(f)$ of successfully completing its execution. If
successfully executed, the function returns a boolean value. We are
interested in the returned value of a function in case it appears as
the rhs of an assignment because the assigned variable may
appear in a condition.  The probability of assigning \emph{true} to
the lhs variable of the statement is the probability that the function
at the rhs returns \emph{true}, which is the product $\mathit{Pr}_S(f)
\cdot \mathit{Pr}_T(f)$, where $\mathit{Pr}_T(f)$ is the
\emph{conditioned} probability that $f$ returns \emph{true} given that
it has been successfully executed. For the sake of readability, we
make the simplifying assumption that all functions whose return value
is used in an assignment are always successful, i.e., have
$\mathit{Pr}_S(f)=1$. Thanks to this assumption the probability of $f$
returning \emph{true} coincides with $\mathit{Pr_T}(f)$ and allows us
to avoid cumbersome, though conceptually simple, formulae in the
following development.

For the conditions $e \in E$  of \textit{if} and 
\textit{while} statements, $\mathit{Pr}_T(e)$ denotes the probability 
of $e$ to be evaluated to \emph{true}.  In case of an \textit{if}
statement, the evaluation of a condition $e$ leads to a probability
$\mathit{Pr}_T(e)$ of following the \textit{then} branch, and
$1-\mathit{Pr}_T(e)$ of following the
\textit{else} branch.  For \textit{while} statements,
$\mathit{Pr}_T(e)$ is the 
probability of executing one iteration of the loop.
  The probability of a condition to be
evaluated to \textit{true} or \textit{false} depends on the current
usage profile and can be estimated on the basis of the designer's
experience, the knowledge of the application domain, or gathered from
previous executions or running instances by combining monitoring and
statistical inference techniques~\cite{fac}.

The value of $\mathit{Pr}_T(e)$ is computed as follows.
If the predicate is the placeholder $\ast$, the probability is
indicated as $\mathit{Pr}_T(\ast)$. If $e$ is a combination of boolean
predicates on variables, the probability value is defined with
respect to its atomic components (assuming probabilistic
independence among the values of the variables in $V$):
\begin{compactitem}[-]
	\item $e=\text{\texttt{"v==true"}} \implies \mathit{Pr}_T(e)=\mathit{Pr}_T(v)$
	\item $e=\text{\texttt{"v==false"}} \implies \mathit{Pr}_T(e)=1-\mathit{Pr}_T(v)$
	\item $e=e_1 \land e_2 \implies \mathit{Pr}_T(e)=\mathit{Pr}_T(e_1) \cdot \mathit{Pr}_T(e_2)$
		\item $e=\neg e_1  \implies \mathit{Pr}_T(e)=1-\mathit{Pr}_T(e_1)$
\end{compactitem}

The initial value of $\mathit{Pr}_T(v)$ for a variable $v \in V$ is
undefined; after the variable is assigned, it is defined as follows:
\begin{compactitem}[-]
	\item \texttt{v:=true} $\implies \mathit{Pr}_T(v)=1$
	\item \texttt{v:=false}  $\implies \mathit{Pr}_T(v)=0$
	\item \texttt{v:=f()} $\implies \mathit{Pr}_T(v)=\mathit{Pr_T}(f)$
\end{compactitem}

The reliability of a program is computed as the \emph{expected
  probability value} of its successful
completion.  To simplify the
mathematical description, we assume independence among all the failure
events.

The reliability of a sequence of statements is essentially the
probability that all of them are executed successfully.  Given the
independence of the failure events, it is the product of the
reliability value of each statement.

For an \textit{if} statement with condition $e$, its reliability is
the reliability of the \textit{then} branch weighted by the
probability of $e$ to be \emph{true}, plus the reliability
of the \textit{else} branch weighted by the probability of $e$ to be
\emph{false}. This intuitive definition is formally grounded on the
law of total probability and the previous assumption of
independence.

The reliability of a \textit{while} statement with condition $e$ and
body $b$ is
determined by the number of iterations $k$. We also assume that
$\mathit{Pr}_T(e)<1$, i.e., there is a non-zero probability of 
exiting the loop, and that $\mathit{Pr}_T(e)$ does not change during
the iterations.
The following formula is easily derived by applying well-known properties of probability theory:
\begin{equation*}
  \begin{split}
    E(\mathit{Pr}_S(\synt{while})) &=
    \sum_{k=0}^{\infty}(\mathit{Pr}_T(e) \cdot
    \mathit{Pr}_S(b))^k \cdot (1-\mathit{Pr}_T(e)) \\
    &= \frac{1-\mathit{Pr}_T(e)}{1-\mathit{Pr}_T(e) \cdot
      \mathit{Pr}_S(b)}
  \end{split}
\end{equation*}
A different construction of this result
can be found in~\cite{Distefano2011}.

We are now ready to encode this analysis through the following attributes:
\begin{compactitem}[-]
\item $\mathit{SYN}(\synt{S}) = \mathit{SYN}(\synt{stmlist}) =
\mathit{SYN}(\synt{stmt}) = \{\gamma,\vartheta\}$;
\item $\mathit{SYN}(\synt{cond}) = \{\delta\}$; 
\item $\mathit{SYN}(\synt{function-id}) = \mathit{SYN}(\synt{var-id}) = \{\eta\}$;
\end{compactitem}
where:
\begin{compactitem}
 \item $\gamma$ represents the reliability of the execution of the
  subtree rooted in the node the attribute corresponds to.
 \item $\vartheta$ represents the knowledge acquired after the execution
  of an assignment. Precisely, $\vartheta$ is a set of pairs 
  $\langle v, \mathit{Pr}_T(v) \rangle$ with $v \in V$ such that there are no two different pairs
  $\langle v_1, \mathit{Pr}_T(v_1) \rangle, \langle v_2, \mathit{Pr}_T(v_2) \rangle \in \vartheta$ with $v_1 = v_2$. If $\nexists \langle v_1, \mathit{Pr}_T(v_1) \rangle \in \vartheta$ no knowledge has been gathered concerning the value of a variable $v_1$. If not differently
  specified, $\vartheta$ is empty.
 \item $\delta$ represents $\mathit{Pr}_T(e)$, with $e$ being the 
  expression associated with the corresponding node.
 \item $\eta$ is a string corresponding to the literal value of an identifier.
\end{compactitem}

The actual value of $\gamma$ in a node has to be evaluated with
respect to the information possibly available in $\vartheta$. For
example, let us assume that for a certain node $n_1$, $\gamma(n_1)=.9 \cdot \mathit{Pr}_T(v)$.
This means that the actual value of $\gamma(n_1)$ depends on the value of the
variable $v$. The latter can be decided only after the execution of an assignment
statement. If such assignment happens at node $n_2$, the attribute $\vartheta(n_2)$ will contain the pair $\langle v, \mathit{Pr}_T(v) \rangle$. For
example, let us assume $\mathit{Pr}_T(v)=.7$; after the assignment, the actual
value of $\gamma(n_1)$ is refined considering the information in $\vartheta(n_2)$,
assuming the numeric value $.63$.
 We use the notation $\gamma(\cdot) \mid \vartheta(\cdot)$ to describe the operation of 
refining the value of $\gamma$ with the information in $\vartheta$. 
Given that $\gamma(\cdot) \mid \emptyset = \gamma(\cdot)$, the operation will be omitted when
$\vartheta(\cdot) = \emptyset$.

The attribute schema is defined as follows:
\begin{compactenum}
	\item
		\begin{grammar}
			<S> ::= `begin' <stmtlist> `end' 
		\end{grammar}
		$\gamma(\synt{S}) \coloneqq \gamma(\synt{stmtlist})$

	\item
		\begin{compactenum}
			\item
				\synt{stmtlist\textsubscript{0}} \; ::= \synt{stmt} `\texttt{;}' \synt{stmtlist\textsubscript{1}}\\ 
				$ \gamma(\synt{stmtlist\textsubscript{0}}) \coloneqq (\gamma(\synt{stmt}) \cdot \gamma(\synt{stmtlist\textsubscript{1}})) \mid \vartheta(\synt{stmt}) $

			\item
				\begin{grammar}
				<stmtlist> ::=  <stmt> `;'
				\end{grammar}
				$\gamma(\synt{stmtlist}) \coloneqq \gamma(\synt{stmt})$

		\end{compactenum}

	\item
		\begin{compactenum}
			\item
				\begin{grammar}
				<stmt> ::=  <function-id> `(' `)'
				\end{grammar}
				$\gamma(\synt{stmt}) \coloneqq \mathit{Pr}_S(f)$\\ with $f \in F$ and $\eta(\synt{function-id})=f$

			\item
				\begin{grammar}
				<stmt> ::=  <var-id> `:=' `true'
				\end{grammar}
				$ \gamma(\synt{stmt}) \coloneqq 1$,
								
				$\vartheta(\synt{stmt}) \coloneqq \{\langle \eta(\synt{var-id}), 1 \rangle \}$

			\item
				\begin{grammar}
				<stmt> ::=  <var-id> `:=' `false'
				\end{grammar}
				$ \gamma(\synt{stmt}) \coloneqq 1$,
				
				$\vartheta(\synt{stmt}) \coloneqq \{ \langle \eta(\synt{var-id}) , 0 \rangle \}$

			\item\label{ref:nf-assignment}
				\begin{grammar}
				<stmt> ::=  <var-id> `=' <function-id> `(' `)'
				\end{grammar}
				$ \gamma(\synt{stmt}) \coloneqq 1$,
				
				$\vartheta(\synt{stmt}) \coloneqq \{\langle \eta(\synt{var-id}) , \mathit{Pr}_T(\eta(\synt{function-id})) \}$ \\with $f \in F$ and $\eta(\synt{function-id})=f$

			\item
				\synt{stmt} \; ::= \lit{if} \synt{cond} \lit{then}
  				\synt{stmlist\textsubscript{0}} \lit{else}
 				\synt{stmlist\textsubscript{1}} \lit{endif}
				\\$\gamma(\synt{stmt}) \coloneqq \gamma(\synt{stmtlist\textsubscript{0}}) \cdot \delta(\synt{cond}) $ 
\\$\text{\hspace{1.7cm}}+ \gamma(\synt{stmtlist\textsubscript{1}}) \cdot (1-\delta(\synt{cond})) $

			\item\label{ref:nf-while}
				\begin{grammar}
				<stmt> ::=  `while' <cond> `do' <stmtlist> `endwhile'
				\end{grammar}
				$\gamma(\synt{stmt}) \coloneqq \dfrac{1-\delta(\synt{cond})}{1-\delta(\synt{cond}) \cdot \gamma(\synt{stmtlist}) }$		

\end{compactenum}

\item
		\synt{cond} \; ::= \ldots\\
		$\delta(\synt{cond}) \coloneqq \mathit{Pr}_T(e)$, with $\eta(\synt{cond})=e$

\end{compactenum}

We now show how to perform probabilistic verification of reliability properties with
\sidecar on the two versions of the example program of
Fig.~\ref{fig:program-versions}. In the steps of attribute synthesis, for
brevity, we use numbers to refer to corresponding nodes in the syntax
tree of Fig.~\ref{fig:trees}.
As for the reliability of the two functions used in the program, we
assume 
$\mathit{Pr}_S(opA)=.97$, $\mathit{Pr}_S(opB)=.99$.

\subsubsection*{Example Program - Version 1}
Given the abstract syntax tree in Fig.~\ref{fig:trees},
evaluation of attributes leads to the following values:
($\eta$ attributes omitted):\\[5pt]
\begin{tabular}{l@{\hskip 5pt}l|l@{\hskip 5pt}l}
$\gamma(2)$ & $\coloneqq .97;$ & $\gamma(18)$ & $\coloneqq \gamma(19);$ \\
$\gamma(6)$ & $\coloneqq 1;$ & $\gamma(11) $ & $\coloneqq .99 \cdot \delta(12)$ \\
$\vartheta(6)$ & $\coloneqq \{\langle x, 1\rangle \};$ &  & \quad $ + .97 \cdot
(1-\delta(12));$\\
$\delta(12)$ & $\coloneqq \mathit{Pr}_T(\text{\texttt{"x==1"}});$  & $\gamma(10)$ & $\coloneqq \gamma(11);$ \\
$\gamma(15)$ & $\coloneqq .99;$ & $\gamma(5)$ & $\coloneqq (\gamma(6) \cdot \gamma(10)) \mid \vartheta(6) =
.99;$\\
$\gamma(14)$ & $\coloneqq \gamma(15);$ & $\gamma(1)$ & $\coloneqq
\gamma(2) \cdot \gamma(5) = .9603;$ \\
$\gamma(19)$ & $\coloneqq .97;$ & $\gamma(0)$ & $\coloneqq \gamma(1) = \ .9603$.  \\
\end{tabular}
\\[5pt]The resulting value for $\gamma(0)$ represents the reliability
of the program, i.e., each execution has a probability equal to
$.9603$ of being successfully executed.

\subsubsection*{Example Program - Version 2}
Version 2 of the example program differs from version 1 only in the
assignment at line~\ref{change}, 
which leads the incremental parser to build 
the subtree  shown in the box of Fig.~\ref{fig:trees}.
Because the matching condition is satisfied, this subtree is hooked
into node 6 of the original tree. Re-computation of the attributes
proceeds upward to the root, leading to the following final values:\\[5pt] 
\begin{tabular}{ll}
$\gamma(6)$ & $\coloneqq 1;$\\
$\vartheta(6)$ & $\coloneqq \{\langle x, 0\rangle\};$\\
$\gamma(5)$ & $\coloneqq (\gamma(6) \cdot \gamma(10)) \mid
\vartheta(6) = .97;$\\
$\gamma(1)$ & $\coloneqq \gamma(2) \cdot \gamma(5) =.9409 $;\\
$\gamma(0)$ & $\coloneqq \gamma(1) \coloneqq .9409 $.\\
\end{tabular}
\\[5pt]
In conclusion, this example shows that \sidecar re-analyzes only a
limited part of the program and re-computes only a small subset of the attributes.

\subsection{Verification of Safety Properties}
\label{sec:reachability-analysis}
This section shows how to use \sidecar to define a basic software
model checking procedure, which solves the safety verification problem:
given a program and a safety property, we want to decide whether there
is an execution of the program that leads to a violation of the
property.

In software model checking, it is common to use a transition-relation
representation of programs~\cite{jhala2009:software-model-}, in which
a program is characterized by a set of (typed) variables, a set of
control locations (including an initial one), and a set of
transitions, from a control location to another one, labeled
with constraints on variables and/or with program operations. Examples of this
kind of representation are control-flow graphs~\cite{aho06:compil} and
control-flow automata~\cite{beyer2007:the-software-mo}. A state of the
program is characterized by a location and by the valuation of the
variables at that location. A computation of the program is a (finite
or infinite) sequence of states, where the sequence is induced by the
transition relation over locations. Checking for a safety property can
be reduced to the problem of checking for the reachability of a
particular location, the \emph{error} location, for example, by
properly instrumenting the program code according to the safety
specification.

In the implementation of safety verification with \sidecar we assume
that the property is defined as a \emph{property
  automaton}~\cite{cheung1999:checking-safety}, whose transitions
correspond either to a procedure call or to a function call that
assigns a value to a variable.  From this automaton we then derive the
corresponding \emph{image automaton}, which traps violation of the
property in an error location (called $\mathit{ERR}$).

\begin{figure}[tb]
\centering
\begin{tikzpicture}[shorten >=1pt,node distance=3cm,auto, initial text=, >=stealth] 
    \node[state, initial] (q0) {$q_0$};
    \node[state] (err) [below right of= q0] {ERR};
    \node[state] (q1) [above right of=err] {$q_1$};
    \path[every node/.style={font=\tt},->] 
    (q0) edge [bend left]  node {opA} (q1)
         edge [dashed, bend right] node {opB} (err)
    (q1) edge [bend left]  node {opB} (q0)
         edge [dashed, bend left]  node {opA} (err)
    ;
  \end{tikzpicture}
\caption{A property automaton; dashed lines belong to the
  corresponding image automaton}
\label{fig:prop1}
\end{figure}
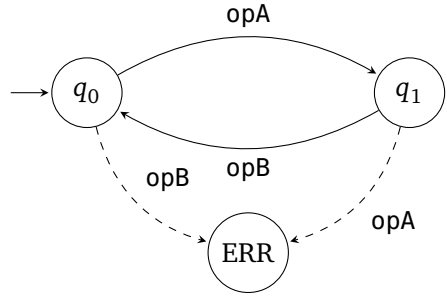

Formally, let $\mathit{VA}$ be the set of variable assignments
from functions, i.e., $\mathit{VA}=\{ x := f \mid x \in V \text{ and }
f \in F \}$.  A property automaton $A$ is a quadruple $A=\langle S, T,
\delta, s_0 \rangle$ where $S$ is a set of locations, $T$ is the alphabet
$T = F \cup \mathit{VA}$, $\delta$ is the transition function $\delta
\colon S \times T \rightarrow S$, and $s_0$ is the initial
location. Given a property automaton $A$, the corresponding image
automaton $A^\prime$ is defined as $A^\prime=\langle S \cup
\{\mathit{ERR}\}, T, \delta^\prime, s_0 \rangle$, where $\delta^\prime
= \delta \cup \{(s,t,\emph{ERR}) \mid (s,t) \in S \times T \land \neg
\exists s^\prime \in S \mid (s,t,s^\prime) \in \delta \}$.  An example
of a property automaton specifying the alternation of operations
\texttt{opA} and \texttt{opB} on sequences starting with \texttt{opA}
is depicted in Fig.~\ref{fig:prop1}; transitions drawn with a dashed
line are added to the property automaton to obtain its image
automaton.

Instead of analyzing the program code instrumented with the safety
specification, we check for the reachability of the error location in an
execution trace of the image automaton, as induced by the syntactic
structure of the program.  

More specifically, each location of the automaton is paired with a
configuration of the program, which consists of a mapping of the
program variables and of the traversal conditions for the paths taken
so far. A configuration is invalid if the set of predicate conditions
holding at a certain location of the program are not compatible with
the current variables mapping for that location. Formally, let
$\mathit{VM}: V \mapsto \{\mathit{true},\mathit{false}\}$ be a mapping
from program variables to their value (if defined).  The set of
possibile configurations that can be reached during the execution of a
program is denoted by $C = (\mathit{VM} \times E) \cup \{ \bot \}$,
where $E$ is the set of
boolean expressions that can appear as the condition of an \emph{if} or a
\emph{while} statement and $\bot$ stands for an invalid
configuration.

Configurations of the program may change when variables are assigned a
new value, e.g., by a direct assignment of a literal or by assigning
the return value of a function. We use a function \upd that updates a
configuration and checks whether it is valid or not.  The function \upd is
defined as $\upd \colon (C \times V \cup \{\varepsilon\} \times
\{\mathit{true},\mathit{false}\} \cup \{\varepsilon\} \times
E \cup \{\varepsilon\} \times
\{\mathit{true},\mathit{false}\} \cup \{ \varepsilon \}) \rightarrow
C$. The function takes a configuration, a variable, its new value, 
a combination of boolean expressions (corresponding to a certain path
condition), its new value, and returns the new configuration;  the $\varepsilon$ symbol
accounts for empty parameters.

We call the pair $\langle$location of the image automaton, 
configuration of the program$\rangle$ an \emph{extended state}. A safety
property represented as an image automaton is violated if it is
possible to reach from the initial extended state another extended
state whose location component is the $\mathit{ERR}$ location. Each
statement in the program defines a transition from one extended state
to another. 

For example, a procedure call determines the location component in an
extended state by following the transition function of the image
automaton corresponding to the call.  An assignment to a variable
updates the program configuration component of an extended state. In
case a variable is assigned the return value of a function invocation,
both components of an extended state are updated.

Conditions in selection and loop statements are evaluated and the
program configuration of the corresponding extended state is updated
accordingly, to keep track of which path conditions have been taken.
For an \textit{if} statement, we keep track of which extended states
could be reachable by executing the statement, considering both the
\textit{then} branch and the \textit{else} branch. For a
\textit{while} statement, we make the common assumption  that a certain constant $K$ is
provided to indicate the number of unrolling passes of the loop. We
then keep track of which extended states could be reachable, both in
case the loop is not executed and in case the loop is executed $K$
times.

The set of attributes is defined as:
\begin{compactitem}[-]
\item $\mathit{SYN}(\synt{S}) = \mathit{SYN}(\synt{stmlist}) =
\mathit{SYN}(\synt{stmt}) = \{\gamma\}$;
\item $\mathit{SYN}(\synt{cond}) = \{\gamma, \nu \}$; 
\item $\mathit{SYN}(\synt{var-id}) = \mathit{SYN}(\synt{function-id}) = \{\eta\}$;
\end{compactitem}
where:
\begin{compactitem}
\item   $\gamma \subseteq S \times C \times S \times C$ is the relation
that defines a transition from one extended state to another one; 
\item $\nu$ is a string corresponding to the literal value of an expression $e \in
E$; 
\item $\eta$ is a string corresponding to the literal value of an
  identifier.
\end{compactitem}
 For the $\gamma$ attribute of \synt{cond} 
 we use the
symbol $\gamma^T$ (respectively $\gamma^F$) to denote the attribute
$\gamma$ evaluated when the condition \synt{cond} is
\textit{true} (respectively, \textit{false}). 
We also define the operation of composing $\gamma$ relations (
the $\circ$ operator) as follows: $\gamma_1 \circ \gamma_2
= \langle s_1, c_1, s_2, c_2 \rangle$ such that there exist $\langle
s_1, c_1, s_i, c_i \rangle \in \gamma_1$ and $\langle s_i, c_i, s_2,
c_2 \rangle \in \gamma_2$.

The attribute schema is defined as follows, where we use the symbols $s, s_1,
s_2$ and $c, c_1, c_2$ to denote generic elements in $S$ and $C$,
respectively.

\begin{compactenum}
\item

\begin{grammar}
<S> ::= `begin' <stmtlist> `end' 
\end{grammar}

$
\gamma(\synt{S}) \coloneqq \gamma(\synt{stmtlist})
$

\item

\begin{compactenum}
\item

\synt{stmtlist\textsubscript{0}} \; ::= \synt{stmt} `\texttt{;}' \synt{stmtlist\textsubscript{1}}

$ \gamma(\synt{stmtlist\textsubscript{0}}) \coloneqq \gamma(\synt{stmt}) \circ \gamma(\synt{stmtlist\textsubscript{1}})$

\item

\begin{grammar}
<stmtlist> ::=  <stmt> `;'
\end{grammar}

$
\gamma(\synt{stmtlist}) \coloneqq \gamma(\synt{stmt})
$

\end{compactenum}

\item

\begin{compactenum}
\item
\begin{grammar}
<stmt> ::=  <function-id> `(' `)'
\end{grammar}
$\gamma(\synt{stmt}) \coloneqq \langle s_1, c,  s_2,
c \rangle $ such that there is $f \in F$ with
$\delta(s_1,f)=s_2$ and $\eta(\synt{function-id})=f$

\item

\begin{grammar}
<stmt> ::=  <var-id> `:=' `true'
\end{grammar}

$ \gamma(\synt{stmt}) \coloneqq \langle s, c_1, s, c_2 \rangle $ with
$c_2 = \upd(c_1, \eta(\synt{var-id}), \mathit{true}, \varepsilon, \varepsilon )$

\item

\begin{grammar}
<stmt> ::=  <var-id> `:=' `false'
\end{grammar}

$ \gamma(\synt{stmt}) \coloneqq \langle s, c_1, s, c_2 \rangle $ with
$c_2 = \upd(c_1, \eta(\synt{var-id}), \mathit{false}, \varepsilon, \varepsilon )$

\item

\begin{grammar}
<stmt> ::=  <var-id> `=' <function-id> `(' `)'
\end{grammar}

$ \gamma(\synt{stmt}) \coloneqq \langle s_1, c_1, s_2,
c_2 \rangle \cup \langle s_1,c_1, s_2,
c_3 \rangle $ such that there is $f \in F$ with
$\delta(s_1, f) = s_2$, $\eta(\synt{function-id})=f$,
$c_2 = \upd(c_1, \eta(\synt{var-id}), \mathit{true}, \varepsilon,
\varepsilon )$, \\and
$c_3 = \upd(c_1, \eta(\synt{var-id}), \mathit{false}, \varepsilon, \varepsilon )$

\item

  \synt{stmt} \; ::= \lit{if} \synt{cond} \lit{then}
  \synt{stmlist\textsubscript{0}} \lit{else}
  \synt{stmlist\textsubscript{1}} \lit{endif}

$ \gamma(\synt{stmt}) \coloneqq  \gamma^T(\synt{cond})
\circ \gamma(\synt{stmtlist\textsubscript{0}})
\cup \gamma^F(\synt{cond}) \circ \gamma(\synt{stmtlist\textsubscript{1}})
$

\item\label{ref:while-rule}

\begin{grammar}
<stmt> ::=  `while' <cond> `do' <stmtlist> `endwhile'
\end{grammar}

$\gamma(\synt{stmt}) \coloneqq 
\gamma^{\mathit{body}} \circ \gamma^F(\synt{cond})$
where $\gamma^{\mathit{body}} = \left(\gamma^T(\synt{cond}) \circ \gamma(\synt{stmtlist})\right)^K$

\end{compactenum}

\item \synt{cond} \; ::= \ldots\\
$\gamma(\synt{cond}) \coloneqq \gamma^T(\synt{cond}) \cup
\gamma^F(\synt{cond})  =$ \\ $\langle s, c_1,
  s,  c_2 \rangle \cup \langle s, c_1, s,  c_3 \rangle$
where $c_2=\upd(c_1,\varepsilon,\varepsilon,\nu(\synt{cond}),
\mathit{true})$
\\and 
$c_3=\upd(c_1,\varepsilon,\varepsilon,\nu(\synt{cond}),
\mathit{false})$ 
 
\end{compactenum}

We now  show how to
perform safety verification with \sidecar on the two versions of the
example program.  For both examples, we consider the safety property
specified with the automaton in Fig.~\ref{fig:prop1}.

\subsubsection*{Example Program - Version 1}
Given the abstract syntax tree depicted in Fig.~\ref{fig:trees},
attributes are synthesized as follows:\\[5pt]
$\gamma(2) \coloneqq \left \{\langle q_0, c, q_1, c \rangle, \langle
  q_1, c, \mathit{ERR}, c\rangle \right \}$;
\\
$\gamma(6) \coloneqq \langle s, c_1, s,
\upd(c_1,\text{\stt{"x"}},\mathit{true},\varepsilon,\varepsilon)
\rangle$;
\\
$\gamma(12) \coloneqq \gamma^T(12) \cup \gamma^F(12) \coloneqq 
\langle
s, c_1, s,
\upd(c_1,\varepsilon,\varepsilon,\text{\stt{"x==true"}},\mathit{true})
\rangle \cup $ \\ $ \qquad \langle s, c_1, s,
\upd(c_1,\varepsilon,\varepsilon,\text{\stt{"x==true"}},\mathit{false})
\rangle $;
\\
$\gamma(15) \coloneqq \left \{\langle q_1, c, q_0, c \rangle, \langle
  q_0, c, \mathit{ERR}, c\rangle \right \}$;
\\
$\gamma(14) \coloneqq \gamma(15)$;
\\
$\gamma(19) \coloneqq \left \{\langle q_0, c, q_1, c \rangle, \langle
  q_1, c, \mathit{ERR}, c\rangle \right \}$;
\\
$\gamma(18) \coloneqq \gamma(19)$;
\\
$\gamma(11) \coloneqq \gamma^T(12) \circ \gamma(14) \cup \gamma^F(12)
\circ \gamma(18) \coloneqq$ \\ $\langle s, c_1, s,
\upd(c_1,\varepsilon,\varepsilon,\text{\stt{"x==true"}},\mathit{true})
\rangle \circ \left \{\langle q_1, c, q_0, c \rangle, \langle q_0, c,
  \mathit{ERR}, c\rangle \right \} \cup $ \\ $\langle s, c_1, s,
\upd(c_1,\varepsilon,\varepsilon,\text{\stt{"x==true"}},\mathit{false})
\rangle \circ \left \{\langle q_0, c, q_1, c \rangle, \langle q_1, c,
  \mathit{ERR}, c\rangle \right \} \coloneqq $ \\
$\{ \langle q_1, c_1,
q_0,
\upd(c_1,\varepsilon,\varepsilon,\text{\stt{"x==true"}},\mathit{true})\rangle,\langle q_0, c_1, \mathit{ERR},
\upd(c_1,\varepsilon,\varepsilon,\text{\stt{"x==true"}},\mathit{true})\rangle,$
\\
$\langle q_0, c_1, q_1,
\upd(c_1,\varepsilon,\varepsilon,\text{\stt{"x==true"}},\mathit{false})\rangle,
\langle q_1, c_1, \mathit{ERR},
\upd(c_1,\varepsilon,\varepsilon,\text{\stt{"x==true"}},\mathit{false})\rangle
\}$;
\\
$\gamma(10) \coloneqq \gamma(11)$;
\\
$\gamma(5) \coloneqq \gamma(6) \circ \gamma(10) \coloneqq \{\langle q_1, c_1, q_0,
\upd(\upd(c_1,\text{\stt{"x"}},\mathit{true},\varepsilon,\varepsilon),\varepsilon,\varepsilon,\text{\stt{"x==true"}},\mathit{true})
\rangle$, \\
$\langle q_0, c_1, \mathit{ERR},
\upd(\upd(c_1,\text{\stt{"x"}},\mathit{true},\varepsilon,\varepsilon),\varepsilon,\varepsilon,\text{\stt{"x==true"}},\mathit{true})
\rangle$, \\
$\langle q_0, c_1, q_1, \bot \rangle$$,\langle q_1, c_1, \mathit{ERR}, \bot \rangle$$\}$.
\\[5pt]The last two tuples of $\gamma(5)$ are discarded because they
contain a $\bot$ configuration. $\bot$ is returned by \upd;
according to its semantics, the evaluation of the condition
\texttt{"x==true"} to \emph{false} is not compatible with the previous
configuration, where \texttt{x} is assigned the value
\emph{true}. Hence, we have:\\[2pt]
 $\gamma(5) \coloneqq \{ \langle q_1, c_1, q_0,
\upd(\upd(c_1,\text{\stt{"x"}},\mathit{true},\varepsilon,\varepsilon),\varepsilon,\varepsilon,\text{\stt{"x==true"}},\mathit{true})
\rangle$,
\\$\langle q_0, c_1, \mathit{ERR},
\upd(\upd(c_1,\text{\stt{"x"}},\mathit{true},\varepsilon,\varepsilon),\varepsilon,\varepsilon,\text{\stt{"x==true"}},\mathit{true})
\rangle \}$;
\\
$\gamma(1) \coloneqq \gamma(2) \circ \gamma(5) \coloneqq \langle q_0, c, q_0,
\upd(\upd(c,\text{\stt{"x"}},\mathit{true},\varepsilon,\varepsilon),\varepsilon,\varepsilon,\text{\stt{"x==true"}},\mathit{true})\rangle$;
\\
$\gamma(0) \coloneqq \gamma(1) \coloneqq \langle q_0, c, q_0,
\upd(\upd(c,\text{\stt{"x"}},\mathit{true},\varepsilon,\varepsilon),\varepsilon,\varepsilon,\text{\stt{"x==true"}},\mathit{true})\rangle$.
\\[2pt]The resulting $\gamma(0)$ shows that the error location is not
reachable from the initial extended state. Therefore we can conclude that the property
will not be violated by any execution of the program.

\subsubsection*{Example Program - Version 2}
The change in
version 2 of the example program affects node 9 of the subtree shown
in the box of Fig.~\ref{fig:trees}. Attribute evaluation proceeds from
node 6 up to the root, as shown below: \\[2pt]$\gamma(6) \coloneqq
\langle s, c_1, s,
\upd(c_1,\text{\stt{"x"}},\mathit{false},\varepsilon,\varepsilon)
\rangle$;\\
$\gamma(5) \coloneqq \gamma(6) \circ \gamma(10) \coloneqq \{\langle q_1, c_1, q_0, \bot \rangle$, $\langle q_0, c_1, \mathit{ERR},
\bot
\rangle$,\\
$\langle q_0, c_1, q_1, \upd(\upd(c_1,\text{\stt{"x"}},\mathit{false},\varepsilon,\varepsilon),\varepsilon,\varepsilon,\text{\stt{"x==true"}},\mathit{false}) \rangle$ \\
$\langle q_1, c_1, \mathit{ERR},
\upd(\upd(c_1,\text{\stt{"x"}},\mathit{false},\varepsilon,\varepsilon),\varepsilon,\varepsilon,\text{\stt{"x==true"}},\mathit{false})
\rangle$\ $\}$. \\[2pt]
The first two tuples of $\gamma(5)$ are
discarded because they contain a $\bot$ configuration. $\bot$ is
returned by \upd; according to its semantics, the evaluation of the
condition \texttt{"x==true"} to \emph{true} is not compatible with the
previous configuration, where \texttt{x} is assigned the value
\emph{false}. Hence, we have:\\[2pt]
 $\gamma(5) \coloneqq \{ \langle q_0, c_1, q_1, \upd(\upd(c_1,\text{\stt{"x"}},\mathit{false},\varepsilon,\varepsilon),\varepsilon,\varepsilon,\text{\stt{"x==true"}},\mathit{false}) \rangle,$ \\
$\langle q_1, c_1, \mathit{ERR},
\upd(\upd(c_1,\text{\stt{"x"}},\mathit{false},\varepsilon,\varepsilon),\varepsilon,\varepsilon,\text{\stt{"x==true"}},\mathit{false})
\rangle\}$;
\\
$\gamma(1) \coloneqq \gamma(2) \circ \gamma(5) \coloneqq \langle q_0, c, \mathit{ERR},
\upd(\upd(c,\text{\stt{"x"}},\mathit{false},\varepsilon,\varepsilon),\varepsilon,\varepsilon,\text{\stt{"x==true"}},\mathit{false})\rangle$;\\
$\gamma(0) \coloneqq \gamma(1) \coloneqq \langle q_0, c, \mathit{ERR},
\upd(\upd(c,\text{\stt{"x"}},\mathit{false},\varepsilon,\varepsilon),\varepsilon,\varepsilon,\text{\stt{"x==true"}},\mathit{false})\rangle$.
\\[2pt]
By looking at $\gamma(0)$, we notice that the error location is
now reachable, which means that version 2 of the program
violates the safety property. 

Note that we reuse results from the analysis of version~1, 
since $\gamma(10)$ and $\gamma(2)$ have not changed. 
In the
analysis of version 2 we processed only 7 tuples of the state space,
compared with the 26 ones processed for version~1.

\section{Discussion}
\label{sec:discussion}

\sidecar introduces a general methodology for the definition of
incremental verification procedures.
It has only two usage requirements:
\begin{inparaenum}[R1)]
 \item the artifact to be verified should have a syntactic structure
   derivable from an OPG;
 \item  the verification procedure has to be formalized as
synthesis of semantic attributes.
\end{inparaenum}

The parsing algorithm used within \sidecar has a temporal complexity
(on average) linear in the size of the modified portion of the syntax
tree. Hence any change in the program has a minimal impact on the
adaptation of the abstract syntax tree too. Semantic incrementality
allows for minimal (re)evaluation of the attributes, by proceeding
along the path from the node corresponding to the change to the root,
whose length is normally logarithmic with respect to the length of the
program. Thus the use of \sidecar may result in a significant
reduction of the re-analysis and semantic re-evaluation steps. The
saving could be very relevant in the case of large programs and rich
and complex attribute schemas.

We emphasize that the two examples showed in the previous section were not designed to be
directly applied to real-world software verification, but to 
give an intuitive glimpse of the generality of the approach.  The generality and
flexibility of OPGs allow for using much richer languages than the
\emph{Mini} example used in this paper. Moreover, attribute
grammars---being Turing
complete---enable formalizing in this framework any algorithmic schema
at any sophistication and complexity level, posing no theoretical
limitation to using \sidecar. For example, more expressive language
constructs and features (like procedure calls, procedures with reference parameters and
side effects, pointers, shared-variable concurrency, non-determinism)
could be accommodated with attribute schemas more complex both in
terms of the attributes definition and in terms of the type (e.g., AGs
with references~\cite{hedin2000:reference-attri} could be useful when the ``semantics'' of a program
element is not confined within its ``syntactic context'').  More 
generally, richer attribute schemas could support both new language
features and different verification algorithms (e.g.,
abstraction-based techniques for the case of verification of safety
properties, or more realistic assumptions on the probabilistic system
behavior for the case of verification of reliability requirements).
In all these scenarios incrementality would be automatically provided
by the framework, without any further effort for the developer.

We acknowledge that some technical issues should be faced when using
\sidecar in non-trivial practical cases. First, existing grammars
could need to be transformed to satisfy requirement R1: the
transformation (especially when automated) might reduce the
readability of the grammar and could impact on the definition of
attribute schemas. Expressing verification procedures as AGs (to
satisfy requirement R2) could be a non-trivial task too: for instance,
developers might simply be not familiar with the programming paradigm
required by AGs; the reuse of known verification algorithms might be
more or less straightforward and/or effective in the context of AG. We
emphasize, however, that such a non-trivial effort is typically done
once for all at design time, possibly in cooperation with 
domain experts. When the system is in
operation, developers should only care about applying the changes and
automatically (and incrementally) verifying their effects.

The generality of the methodology advocated by \sidecar widens the
scope of application to a number of scenarios.  For example, at design
time, \sidecar (possibly integrated within IDE tools) can effectively
support designers in evaluating the impact of changes in their
products, in activities such as what-if analysis and regression
verification, very common in agile development processes.  Existing
techniques for automated verification based either on model checking
or on deductive approaches, as well as their optimizations, could be
adapted to use \sidecar, exploiting the benefits of incrementality.
At run time, the incrementality provided by \sidecar could be the key
factor for efficient online verification of continuously changing
situations, which could then trigger and drive the adaptation of
self-adaptive systems~\cite{calinescu12:self}. As another instance of the approach's
generality, similarly to the
probabilistic verification described in
section~\ref{secReliabilityAnalysis}, other quantitative properties,
such as execution time and energy consumption, could be verified with
\sidecar. Furthermore, \sidecar could also bring at run time the same
analysis techniques so far limited to design time because of efficiency reasons.

\section{Related Work}
\label{secRelatedWork}

In this section we present related work in two parts. First, we
discuss work that addresses incrementality in verification\footnote{Incidentally, the use of the term
\textit{incremental model checking} in the context of bounded model
checking~\cite{biere2003:bounded-model-c} has a different meaning,
since it refers to the possibility of changing the bound of the
checking.} in general; next, we
discuss other incremental approaches in the fields of the two examples
presented here, namely probabilistic verification and safety
program verification.

Different methodologies have been proposed in the literature as the basis
for incremental verification techniques.  They are mainly
grounded in the assume-guarantee~\cite{Jones1983} paradigm.  This
paradigm views systems as a collection of cooperating modules, each of
which has to guarantee certain properties.  The verification methods
based on this paradigm are said to be compositional, since they allow 
reasoning about each module separately and deducing properties about their 
integration.  If the effect of a change can be localized
inside the boundary of a module, the other modules are not affected,
and their verification does not need to be redone. This feature is for
example exploited in~\cite{Cobleigh2003}, which proposes a
framework for performing assume-guarantee reasoning in an incremental
and fully automatic fashion. Assume-guarantee based verification has
been exploited also for probabilistic reasoning (e.g.,
in~\cite{Kwiatkowska2010tacas}), even though we are not aware of approaches
using it in an incremental fashion.

Focusing now, more specifically, on incremental probabilistic
verification, a known technique to achieve incremental verification is
\textit{parametric analysis}~\cite{daws2005symbolic}.  With this
technique, the probability values of the transitions in the model that
are supposed to change are labeled with symbolic parameters. The model
is then verified providing results in the form of closed mathematical
formulae depending on the symbolic parameters.  As the actual values
for the parameters become available (e.g., during the execution of the
system), they are replaced in the formulae, providing a numerical
estimation of the desired reliability. Whenever there is a change of
the values of the parameters, the results of the preprocessing phase
can be reused, with significant improvements of the verification
time~\cite{Filieri2011a}.  The main limitation of this approach is
that a structural change in the software invalidates the results of
the preprocessing phase, requiring the verification to start from
scratch, with consequent degradation of the analysis performance.

Parametric analysis is reminiscent
of the notion of \emph{partial evaluation}, originally introduced
in~\cite{Ershov1977}. Partial evaluation can be seen as a
transformation from the original version of the program to a new
version called \emph{residual program}, where the properties of
interest have been partially computed against the static parts,
preserving the dependency on the variable ones. As soon as a change is
observed, the computation can be moved a further step toward
completion by fixing one or more variable parts according to the
observations.

Concerning related work on incremental safety verification, other
approaches based on (regression) model checking reason in terms of the
representation (e.g., a state-transition system) explored during the
verification, by assessing how it is affected by changes in the
program. The main idea is to maximize the reuse of the state space
already explored for previous versions of the program, isolating the
parts of the state space that have changed in the new version.  The
first work in this line of research addressed modal
mu-calculus~\cite{sokolsky1994:incremental-mod}.  Henzinger et
al.~\cite{henzinger2004:extreme-model-c} analyze a new version of the
program by checking for the conformance of its (abstract) state space
representation with respect to the one of the previous version. When a
discrepancy is found, the algorithm that recomputes the abstraction is
restarted from that location. Depending on where the change is
localized in the program text, the algorithm could invalidate---and
thus recompute---a possibly large portion of the program state space.
Similarly, incremental approaches for explicit-state model checking of
object-oriented programs,
such as~\cite{lauterburg2008:incremental-sta}
and~\cite{yang2009:regression-mode}, analyze the state space checked
for a previous version and assess, respectively, either the transitions that  
do not need to be re-executed in a certain exploration of the state
space,
 or the states that can be pruned, because not affected by the code
 change.
These approaches tie incrementality to the low-level details of the
verification procedure, while \sidecar supports incrementality at a higher
level, independently on the algorithm and data structures defined in the attributes.
Conway et
al.~\cite{conway2005:incremental-alg} define incremental algorithms
for automaton-based safety program analyses. Their granularity for the
identification of reusable parts of the state space is coarse-grained,
since they take a function as the unit of change, while \sidecar has a
finer granularity, at the statement level.  A combination of a modular
verification technique that also reuse cached information from the
checks of previous versions is presented
in~\cite{krishnamurthi2007:foundations-of-} for aspect-oriented
software.

In conclusion, the syntactic-semantic approach embedded in \sidecar 
 does not constrain incrementality depending on
on the modular structure of the artifacts, as instead required by
assume-guarantee approaches. Furthermore, it provides 
a general and unifying methodology for defining verification
procedures 
for functional and non-functional requirements.

\section{Conclusion and Future Work}
\label{secConclusions}

Incrementality is one of the most promising means to dealing with
software evolution.  In this paper we addressed the issue of
incrementality in verification activities by introducing \sidecar, a
framework for the definition of verification procedures, which are
automatically enhanced with incrementality by the framework itself.  \sidecar supports a
verification procedure encoded as synthesis of semantic attributes
associated with a grammar. The attributes are evaluated by traversing
the syntax tree that reflects the structure of the software system. By
exploiting incremental parsing and attributes evaluation techniques,
\sidecar reduces the complexity of the verification procedure in
presence of changes.
We have shown \sidecar in use to define two
kinds of verification, namely probabilistic verification of
reliability properties and safety verification of programs.

Future work will address several directions. We want to
support run-time changes of the language (and thus the grammar) in
which the artifact to be verified is described, motivated by advanced
adaptiveness capability scenarios. We also want to support
changes in the properties to be verified, and still exploit the
benefit of incremental verification. We will continue our work
to develop an incremental verification environment---by incorporating
improvements to exploit parallelism~\cite{barenghi12} and to apply
finer incremental parsing techniques---and will conduct experimental
studies on real-world applications to quantify the effectiveness of
\sidecar in the definition and the execution of state-of-the-art
verification procedures, identifying the kind of verification
procedures for which the \sidecar approach is more cost-effective. We will
also investigate the pragmatic issues discussed in
section~\ref{sec:discussion}, i.e., what can be reasonably encoded
using OPGs and AGs. Finally, we plan to exploit \sidecar to introduce
verification-driven development in iterative and/or agile development processes.

\section*{Acknowledgments}
This work has been partially supported by the European Community under
the IDEAS-ERC grant agreement no.~227977-SMScom and by the National
Research Fund, Luxembourg (FNR/P10/03).
\bibliographystyle{abbrv}
\bibliography{new-incremental-tr}
\end{document}